\documentclass{article}
\usepackage[utf8]{inputenc}
\usepackage{amsmath}
\usepackage{graphicx}
\usepackage{subcaption}

\usepackage{lineno}

\title{Elastic 3D Wavefield Simulation on budget GPUs using the GLSL shading language}
\author{Emanuel Trabes 
Silvana Spagnotto
$^2$ $^3$, \\
Orlando Alvarez Pontoriero
$^3$ $^4$,
Julio Daniel Dondo Gazzano 
$^1$ \\
and Carlos Federico Sosa Paez 
$^1$}

\date{%
    $^1$Departamento de Electronica, Facultad de Ciencias Fisico Matematicas y Naturales, Universidad Nacional de San Luis (UNSL), Argentina\\%
    $^2$Facultad de Ciencias Fisico Matematicas y Naturales, Universidad Nacional de San Luis (UNSL), Argentina\\[2ex]%
    $^3$Consejo Nacional de Investigaciones Científicas y Técnicas (CONICET), Argentina\\%
    $^4$ Instituto Sismologico Ing. F.S.Volponi, Facultade De Cs. Exactas Fisicas y Naturales, Universidad Nacional de San Juan (UNSJ), Argentina\\[2ex]%
}

\begin{document}

\maketitle

\begin{abstract}
Forward wavefield simulation is an important step in Full Waveform Inversion systems. Fast simulations are instrumental to get inversion result in reasonable time frames. Most of research and software aims towards utilizing costly computer clusters composed of multiple CPUs and numerous high end GPUs to shorten the forward simulation time. Using this type of hardware has some disadvantages as: high cost, complex programming models and unavailability of resources. In this work, we present a finite difference elastic 3D wavefield forward simulation that takes advantage of any modern low end GPU, by using the GLSL shading language.
Some of the advantages of using GLSL are: runs in any modern GPU, has a simplified computing and memory model and provides state of art performance thanks to its very well optimized vendor developed drivers. We show that our GLSL implementation easily outperforms a multicore CPU implementation in a modern PC. We further benchmark our result using a real seismic event, and show that we can get accurate simulations in reasonable time using our system. 

\end{abstract}

\section{Introduction}

Finite difference (FD) methods are some of the most flexible techniques for elastic wavefield simulation. These methods easily simulate wavefields of highly heterogeneous mediums with complex surface profiles. At the same time they can reproduce highly accurate full waveforms of simulated seismograms  \cite{paper_virieux_1} \cite{paper_virieux_2} \cite{paper_finite_difference}. As forward simulations are a necessary component of any Full Waveform Inversion (FWI) method, developing fast and accurate forward simulations is a key issue in this context.

Classical forward waveform simulations are based on frequency techniques \cite{paper_ISOLA}\cite{paper_ISOLA_2}, and most of them are only able to handle 1D velocity models \cite{review_de_metodos}. When the medium is highly heterogeneous or when the size of the zone under study is large, 1D velocity models do not faithfully represent the medium. 
Because low frequencies are less prone to being affected by low quality velocity models, classical methods utilize only the low frequency components of the full seismogram records \cite{kikuchi_1}\cite{kikuchi_2}\cite{Herrmann}. Consequently, FWI methods that utilize frequency techniques to accomplish the forward simulation are only able to recover very coarse medium parameters. To recover small details in the medium, methods like FD are better suited. 


One of the main reasons why finite difference methods are not currently popular among full waveform inversion methods is its high computational cost. Traditionally this problem would be resolved by using a computer cluster, which is accessed by the user remotely \cite{Simulacion_usando_GPU_1}. Naturally, there are complications with this solution, namely with the cost of the computer cluster and its maintenance, having to schedule computing time beforehand, and the complex programming that involves using such hardware, like load distribution and memory management.
This suggests that a more flexible solution should be explored. An interesting alternative could be the use of Graphic Processing Units (GPUs).

Most of the GPU implementations of finite difference methods available in the literature use the CUDA Application Programming Interface (API) \cite{Simulacion_usando_GPU_2} \cite{Simulacion_usando_GPU_3} \cite{Simulacion_usando_GPU_4} \cite{Simulacion_usando_GPU_1}. This API requires the use of a somewhat complicated programming model. The programmer must explicitly distribute the computing task among the GPU’s Processing Units (PU), do explicit memory management both between PC and GPU memory and among the different types of memory available on the same GPU. The PU workload and memory management methodology vary between GPUs of different manufacturers, and also among GPUs of the same manufacturer within different GPUs generations. Also, this API is only available in some specialized GPUs manufactured by NVIDIA.

Instead of using CUDA, an alternative is to use the GLSL shading language provided by the OpenGL API. This API is specifically developed for 3D graphic workloads. Because finite difference methods share most of its computing requirements with graphic workloads, OpenGL can be effectively used for wavefield simulation.
OpenGl doesn't need explicit memory management nor manual workload distribution as CUDA.
Also, as OpenGL is widely used for 3D graphics, the drivers provided by the GPU manufacturer are continuously optimized with the latest updates and achieves the maximum performance possible by the hardware. In the case of CUDA, the manufacturer often drops support for old GPU generations or does not provide sufficiently optimized drivers. OpenGL is supported by virtually all GPUs manufacturers, and thus the same software written with this API can run in PCs, notebooks and even in smartphones. This ubiquitous availability of OpenGL would allow the user to run simulations and experiments even when he is working in situ in a zone where there is no computer clusters or internet availability. 

In this work, we used the OpenGL API to accomplish finite difference elastic wave simulation in low cost GPUs. One interesting realization is that finite difference methods share most of the computing tasks with traditional 3D graphics pipelines. For this reason, we used the OpenGL API to develop a finite difference staggered grid elastic wavefield simulator. We fully exploited the benefits of the OpenGL API, which allowed us to write highly readable and customizable code, without cumbersome specialized algorithms dealing with memory management and workload distribution. We further used the API to accomplish highly detailed trilinear interpolation of the data, which allowed us to easily get effective medium parameters as well as to read the simulated seismogram recording in sub-staggered grid positions. All the system is written in the GLSL 3.3 shading language, and the computation is started from a python script, which allows to easily integrate our simulation framework with modern seismograph processing libraries such as obspy. 
We compared our system with a multicore CPU implementation, and finalized this work comparing our simulations with real seismograms recorded from a 5.2 magnitude earthquake originated in Cuba in 2016.


\section{GLSL implementation}

\subsection{Elastic wave equations}

The equations that describe elastic wave propagation can be expressed as two coupled first order differential equations, as described in \cite{paper_finite_difference} 

\begin{equation}
\begin{gathered}
\rho(p) \dot{v}(p,t) = \nabla \sigma(p,t) + f(p,t)
\label{equ:velocityEquation}
\end{gathered}
\end{equation}

\begin{equation}
\begin{gathered}
\dot{\sigma}(p,t) = C(p) \cdot (\frac{1}{2}(\nabla v(p,t) + (\nabla v(p,t))^T))
\label{equ:stressEquation}
\end{gathered}
\end{equation}

Where $v = [v_x, v_y, v_z]^T$ is the velocity vector of a point in the medium at space coordinate $p = [x,y,z]^T$ and time $t$. The symmetric stress tensor 
$\sigma =$

\begin{equation}
\begin{bmatrix}
\sigma_{xx} & \sigma_{xy} & \sigma_{xz}\\
\sigma_{xy} & \sigma_{yy} & \sigma_{yz}\\
\sigma_{xz} & \sigma_{yz} & \sigma_{zz}
\end{bmatrix}
\end{equation}

is linearly related to the strain tensor $\varepsilon = \frac{1}{2}(\nabla v(p,t) + (\nabla v(p,t))^T) $ by the stiffness tensor $C$. In the case of an isotropic medium, this tensor takes a special form where only two degrees of freedom remain: the Lamé parameters $\lambda$ and $\mu$. Together with the density $\rho$, these 3 parameters define the medium characteristics. These differential equations can be trivially parallelized, discretizing both the position $p$ and time $t$, and then computing the approximated spatial and temporal derivatives. 

\subsubsection{Staggered Grid}
To avoid the common checkerboard pattern problem present in coalesced grid implementations \cite{paper_virieux_1}, our system follows a staggered grid formulation, following the work of \cite{Finite_difference_graves}. In this implementation, $\sigma_{xx}$, $\sigma_{yy}$ and $\sigma_{zz}$ are defined at positions $[x,y,z]$, $\sigma_{xy}$ at $[x+0.5,y+0.5,z]$, $\sigma_{xz}$ at $[x+0.5,y,z+0.5]$ and $\sigma_{yz}$ at $[x,y+0.5,z+0.5]$. At the same time, $v_x$ is defined at $[x+0.5,y,z]$, $v_y$ at $[x,y+0.5,z]$ and $v_z$ at $[x,y,z+0.5]$

\subsubsection{4th Order Approximation}

To approximate $\nabla \sigma(p,t)$ and $\nabla v(p,t)$, it is common in the literature to use a 4th order difference approximation at the different staggered grip positions \cite{paper_virieux_1} \cite{Finite_difference_graves}.

\begin{equation}
\frac{\partial f(x)}{\partial x} = -\frac{1}{24}f(x+1.5) + \frac{9}{8}f(x+0.5) - \frac{9}{8}f(x-0.5) + \frac{1}{24}f(x-1.5) 
\label{equ:derivativeAproximation}
\end{equation}

\subsubsection{Source frequency and Simulation time step}

As noted in \cite{paper_finite_difference}, the maximum allowable simulation time step responds to

\begin{equation}
dt_{max} = 0.495 \cdot \frac{dx_{min}}{vel_{max}} 
\label{equ:max_dt}
\end{equation}

and the maximum frequency:

\begin{equation}
frec_{max} = \frac{vel_{min}}{5.0*dx_{max}}
\label{equ:max_frec}
\end{equation}

where $dx_{max}$ $dx_{min}$ is the maximum/minimum space step in the $x$ $y$ or $z$ direction and $vel_{max}$ and $vel_{min}$ correspond to the maximum and minimum medium velocity.

\subsection{Memory Management}
\label{sec:MemoryManagment}

The simulation described in equations (\ref{equ:velocityEquation}), (\ref{equ:stressEquation}) and (\ref{equ:derivativeAproximation}) requires to execute $4$ memory access for every derivative calculation, witch add to $9$ derivatives for (\ref{equ:velocityEquation}) and $9$ for (\ref{equ:stressEquation}). Simultaneously, to accomplish good accuracy, we need to solve (\ref{equ:velocityEquation}) and (\ref{equ:stressEquation}) in thousands of grid points, which utilizes large amounts of memory resources. As a consequence, the forward simulation algorithm is heavily memory bound \cite{Simulacion_usando_GPU_1}. To accomplish low simulation times, good memory allocation and management is crucial.

Using the OpenGL API, we have defined a floating point 3D texture for every variable in the simulation, namely the velocity components $v_x$, $v_y$ and $v_z$, the stress tensor components $\sigma_{xx}$, $\sigma_{yy}$, $\sigma_{zz}$, $\sigma_{xy}$, $\sigma_{xz}$, $\sigma_{yz}$ and the medium parameters $\rho$ $\lambda$ and $\mu$. 


Due to the fact that the differential equation for $\dot{v}$ do not depend directly on $v$, nor the equation for $\dot{\sigma}$ depends directly on $\sigma$, we can solve (\ref{equ:velocityEquation}) and (\ref{equ:stressEquation}) in place by using OpenGL blend equations. The blend functions allow to combine the result of every fragment output with the previous data contained in the framebuffer. We set the framebuffer to be the $v_x$ $v_y$ and $v_z$ textures, and then compute the right side of equation (\ref{equ:velocityEquation}). This operation adds the previous value of $v_x$ $v_y$ and $v_z$ with the update defined in the right side of (\ref{equ:velocityEquation}), effectively solving for $\dot{v}$. We do the same to solve (\ref{equ:stressEquation}). As a consequence, we just need to store in memory one copy of each $v$ and each $\sigma$, saving on memory resources.

There is no need to explicitly perform memory management in the shader, the OpenGL driver manages memory automatically. The high locality of the algorithms data access and its similarity with graphic workloads makes it easy for OpenGL to optimize memory caches.


\subsubsection{Trilinear interpolated effective medium parameters}



The staggered grid implementation needs a method to evaluate the medium parameters $\rho$ at the location of each $v$, and $\lambda$ and $\mu$ at the locations of each $\sigma$, which due to the staggered grid are not located at coalesced memory positions. There is no need to explicitly calculate mean values at non-integer positions, as made by \cite{paper_virieux_1} \cite{Finite_difference_graves}, GLSL allows to get trilinear interpolated values automatically at runtime in any real value position. We defined the OpenGL texture for $\rho$, $\lambda$ and $\mu$ with a linear interpolation, both for the maximization and minimization of the texture. Then the texture read is performed in the shader, and the OpenGL driver automatically interpolates the values. 
Using this functionality of OpenGL, we can define sizes for the medium parameter textures independently from the simulation grid size.
As in most cases only a very coarse medium model is available, we can define the textures for $\rho$ $\lambda$ and $\mu$ to be significantly smaller in size than the textures for $v$ and $\sigma$. Then we can get a value of $\rho$ $\lambda$ and $\sigma$ at a particular position $p$ by the automatic trilinear interpolation provided by OpenGL.


\subsubsection{Out of bound memory fetches}

The computation of (\ref{equ:derivativeAproximation}) at the border should be specially considered, as at position $x = 0$ there is no data at $x - 1.5$. OpenGL allow us to solve this issue without the need to code special conditions in the shaders. We simply configure the OpenGL texture for all the variables to do a mirrored repeat texture wrapping in all off its $3$ axis. This way when shader tries to access a out-of-bound memory location, OpenGL fetches it its mirrored repeated position, meaning that if $x=0$ and the shader tries to access position $x - 1.5$, OpenGL fetches data from $x + 1.5$. Because the edges are in zones where data should be dampen, there is no degradation on the simulation precision. 

\subsection{Shader design and workload distribution}

Due to OpenGLs simplified computation model, the shaders used are rather simple. In 2 main shaders we computed the velocities $v_x$ $v_y$ and $v_z$ following equations (\ref{equ:velocityEquation}) and  (\ref{equ:derivativeAproximation}), and $\sigma_{xx}$ $\sigma_{yy}$ $\sigma_{zz}$ $\sigma_{xy}$ $\sigma_{xz}$ and $\sigma_{yz}$ following equations (\ref{equ:stressEquation}) and  (\ref{equ:derivativeAproximation}). In each shader there is no special code for out of bound fetches, as explained before. Also there is no special code for workload distribution, OpenGL distributes the computing among the available GPUs PU automatically. As we mentioned before, the similarities of wave simulation with graphic workloads makes it easy for the OpenGL driver to distribute the computing workload efficiently. 

Commonly, OpenGL is used to render to 2D textures. To render to a 3D texture with just one CPU render call, we utilized \emph{instanced rendering}. We designed the shaders so that each time only one $z$ grid step is computed. Then, using \emph{glDrawArraysInstanced} function, we command OpenGL to render each different $z$ grid step in one single rendering call. 

\subsection{Retrieving simulated receivers data}

Exchanging data between CPU main memory and GPU memory can be a slow process, and it should be avoided as much as possible. For this reason, we dont retrieve the simulated receivers data at every time step. Instead, we defined a 2D texture in the GPU, where we store every $v_x$ $v_y$ and $v_z$ at all receivers locations at every time step in the simulation. When the simulation is over we retrieve all data, so reading the data from the GPU to the CPU does not have a negative impact in the simulation time. 

Furthermore, we take advantage of the trilinear interpolation of the data available in OpenGL. We recover $v_x$ $v_y$ and $v_z$ at the station location $p_l$ by trilinearly interpolating velocities at grid positions $p$.

\subsubsection{Obspy interoperability}

We save the receiver data in obspy’s trace/stream data classes. Supporting popular libraries like obspy make the code more useful and easier to use. All of the processing classes available in obspy are accessible to be used in the resulting simulation data.


\subsection{Boundary conditions}

There are two boundary conditions that must be taken into account. The first is due to the limited grid size, where waves traveling towards the simulation limits should not bounce back. The second condition arises while simulating the earth surface. For the first boundary condition, we utilize a dampening factor in the edges of the simulation, so the waves going outwards are progressively dampened. 
For the surface boundary condition we utilized the so called \emph{vaccum formulation} \cite{Finite_difference_graves}, where the values for $\rho$, $\lambda$ and $\mu$ and are set to low values above the surface, instead of
using an involved solutions like those developed in \cite{superficie_1} \cite{superficie_2} \cite{superficie_3}.
It is known that this formulation generates inaccuracies in the simulation when using a 4th order derivative approximation like (\ref{equ:derivativeAproximation}) \cite{Finite_difference_graves}. To avoid inaccuracies, near the surface position we change the derivative approximation (\ref{equ:derivativeAproximation}) to a second order one. 






\section{Case Study}

To evaluate both the precision and the computation time of our simulation, we designed the following experiment. We utilized as input for our simulation a known seismic source and known medium parameters data, and then compared the resulting simulated seismograms against actual seismogram recordings. The seismogram data utilized was taken from a local network located in Cuba, and its station recordings were provided by the Data Management Center of the Incorporated Research Institutions for Seismology (IRIS-DMC). The earthquake had its origin time on 17 January of 2016, and due to the dense seismograph network, its source position and moment tensor was accurately estimated. To recover the seismic moment tensor we utilized the software package ISOLA, and the focal mechanism published by the PDE-USGS catalog \cite{paper_ISOLA} \cite{paper_ISOLA_2}. To accomplish the simulation we also have at our disposal the instrument response for every station utilized, as well as a 1D velocity model of the area, provided by \cite{medium_params}. 

\subsection{The Mw 5.2, 2016/01/17 Cuban Earthquake and tectonic summary}

Four large tectonic plates interact in this area (the North American plate, the South American plate, the Nazca plate and the Cocos plate), causing a fairly complex tectonic regime. As consequence, this zone is prone to the occurrence of large earthquakes.

\begin{figure}[h]
     \centering  
     \includegraphics[width=0.99\textwidth]{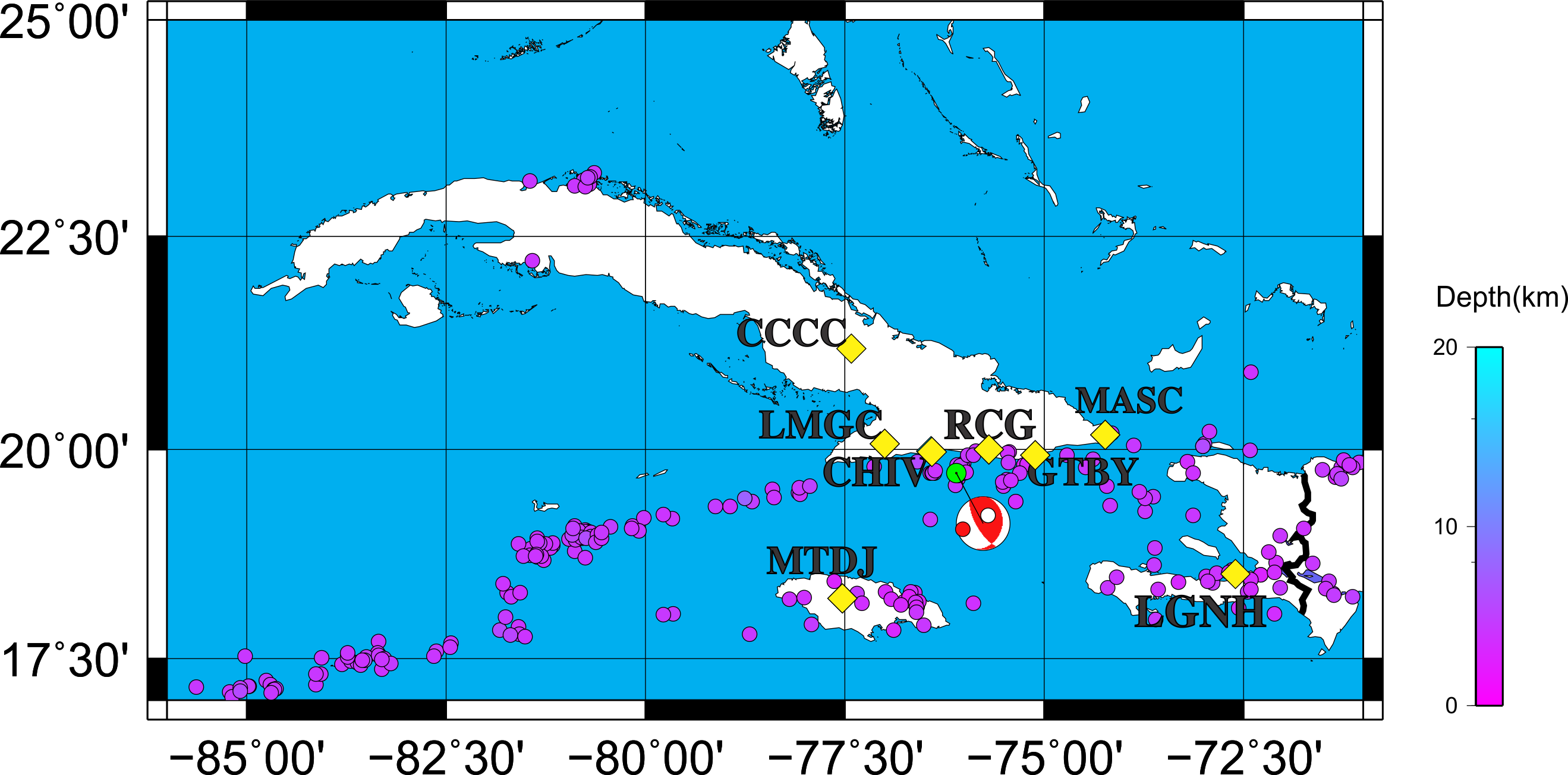}
     \caption{Mw 5.2, 2016/01/17 earthquake and 10 years of seismicity in the region, according to the PDE-USGS catalog. The Focal mechanism is shown using a red beach ball (lower hemisphere compression quadrants), and was acquired from the same catalog. The seismological stations utilized in this experiment are illustrated using yellow diamonds}
     \label{fig:zone_summary}
\end{figure}

The Mw 5.2, 2016/01/17 earthquake occurred at 08:30:23 (UTC) and its epicenter was located at 19.720°N 76.097°W and at 10.0 km depth, according to the PDE-USGS catalog. This earthquake had at least four important foreshocks with magnitude greater than 4, and some aftershocks of moderate magnitude. All events had similar focal mechanisms. In figure \ref{fig:zone_summary}, we illustrate the location and focal mechanism of this event, as well as the epicenter of several earthquakes that occurred in the last 10 years (taken from the PDE-USGS catalog). 

\subsubsection{Station network and earthquake}

We list the location of the seismological stations utilized for this case study in table \ref{table:stations}. Centroid location and time of the earthquake is shown in table \ref{table:earthquake}. We obtained the earthquakes moment tensor by performing an inversion using the ISOLA software package.


\begin{table}[ht]
\centering
\caption{List of stations}
\begin{tabular}{ |c|c|c|c| } 
 \hline
 Station name & latitude & longitude & altitude \\ 
 \hline
 CHIV & 19.9763  & -76.4147  &  20.0  \\
 RCC  & 19.9942  & -75.6958  & 100.0  \\
 LMGC & 20.064   & -77.005   & 167.0  \\
 GTBY & 19.92681 & -75.11081 &  79.2  \\
 MASC & 20.175   & -74.231   & 350.0  \\
 CCCC & 21.1934  & -77.4173  &  89.55 \\
 MTDJ & 18.22606 & -77.53453 & 925.0  \\
 LGNH & 18.511   & -72.6058  &  62.0  \\
 \hline
\end{tabular}
\label{table:stations}
\end{table}%

\begin{table}[ht]
\centering
\caption{Details of the earthquake}
\begin{tabular}{ |c|c|c| } 
 \hline
Origin time & 2016/1/17  8:30:23 \\
 \hline
Centroid time & 2016/1/17  8:30:25.08 \\
 \hline
Centroid location &  19.749 -76.09  7.0 \\
 \hline
\end{tabular}
\label{table:earthquake}
\end{table}

The estimated moment tensor is detailed in \ref{equ:momentTensor}.

\begin{gather}
  \begin{bmatrix}
 -2.37 \cdot 10^{14} & -3.39 \cdot 10^{15} & -7.79 \cdot 10^{14} \\
 -3.39 \cdot 10^{15} & -4.31 \cdot 10^{16} &  4.60 \cdot 10^{15} \\
 -7.79 \cdot 10^{14} &  4.60 \cdot 10^{15} &  4.33 \cdot 10^{16} \\
   \end{bmatrix}
   \frac{kg \cdot m}{m^2}
   \label{equ:momentTensor}
\end{gather}

\subsubsection{Medium parameters}

The medium parameters were taken from \cite{medium_params}. 
Our simulation method performs calculations in SI, so all parameters are converted accordingly. We calculated the values for the different density layers following the work of \cite{relacion_rho}. 

\begin{table}[ht]
\centering
\caption{Medium parameters}
\begin{tabular}[t]{|c|c|c|c|c|c|}
 \hline
    depth of layer (top)(km)  &   $V_p(km/s)$ &  $V_s(km/s)$ & $\rho(g/cm^3)$ \\
     \hline
      0.0                   &  4.90      &  2.816    &    2.50 \\
      3.0                   &  5.40      &  3.103    &    2.60 \\
      5.0                   &  6.00      &  3.448    &    2.70 \\
      7.0                   &  6.90      &  3.966    &    2.80 \\
     20.0                   &  7.60      &  4.368    &    3.10 \\
     26.0                   &  7.80      &  4.483    &    3.26 \\
     34.0                   &  8.00      &  4.598    &    3.30 \\
      \hline
\end{tabular}
\label{tab:medium_params}
\end{table}%

\subsection{Simulation description}


For this particular station network and earthquake, we configured our simulation with the parameters shown in figure \ref{tab:simulation_limits}.

\begin{table}[ht]
\centering
\caption{Simulation limits}
\begin{tabular}[t]{|c|c|}
\hline
latitude range: & from 17.78 to 21.63 degree \\
\hline
longitude range: & from -78.27 to 71.86 degree \\
\hline 
depth range: & from -30.0 to 90.0 km\\
\hline
Size (km) & x: 427.03 y: 679.36 z: 120.0\\
\hline
Simulation start time & 2016-01-17 08:29:25.0 \\ 
\hline
Simulation end time & 2016-01-17 08:32:24.91 \\
\hline
\end{tabular}
\label{tab:simulation_limits}
\end{table}



\subsubsection{Simulation sizes}

We tested our system with different simulation sizes. At lower simulation sizes, we can have lower computation workloads and greater simulation time steps, as noted in (\ref{equ:max_dt}). But at the same time, the simulation would only allow lower maximum source frequencies, responding to (\ref{equ:max_frec}). The simulation step then controls the level of detail of the simulation, where coarser levels complete faster, but with lower detail. 
The following table summarizes the different level of detail tested in this experiment

\begin{table}[ht]
\centering
\caption{Simulation details}
\begin{tabular}[t]{|c|c|c|c|c|c|c|}
\hline
level & Grid size & time steps &  dp (km) & dt (s) & max f (Hz) \\
\hline
0 & 64x64x32 & 1700 & 6.67x10.61x3.75 & 0.1 & 0.037  \\
\hline
1 & 64x64x64 &  1700 & 6.67x10.61x1.87 & 0.1 & 0.037 \\
\hline
2 & 128x64x64 & 1700 & 3.33x10.61x1.87 & 0.1 & 0.037 \\
\hline 
3 & 128x128x64 & 1700 & 3.33x5.30x1.87 & 0.1 & 0.075 \\
\hline 
4 & 128x128x128 & 3400 & 3.33x5.30x0.93 & 0.05 & 0.075 \\
\hline 
5 & 256x128x128 & 3400 & 1.66x5.30x0.93 & 0.05 & 0.075 \\
\hline
6 & 256x256x128 & 3400 & 1.66x2.65x0.93 &  0.05 & 0.15 \\
\hline
7 & 256x256x256 & 17000 & 1.66x2.65x0.46 & 0.01 & 0.15 \\
\hline
8 & 512x256x256 & 17000 & 0.83x2.65x0.46 & 0.01 & 0.15 \\
\hline
9 & 512x512x256 & 17000 & 0.83x1.32x0.46 & 0.01 & 0.30 \\
\hline
10 & 512x512x512 & 17000 & 0.83x1.32x0.23 & 0.01 & 0.30 \\
\hline
\end{tabular}
\label{tab:lvl}
\end{table}

The maximum grid size used is 512x512x512, to allow all the data to fit inside the GPU memory and thus avoid CPU-GPU memory allocations.

\subsection{CPU implementation}

We compare our GPU results against an equivalent implementation programmed using highly optimized vector algebra package openblas, using the library numpy in a python script.

\subsection{Results}

\begin{figure}[h]
     \centering  
     \begin{subfigure}[b]{0.24\textwidth}
         \centering
         \includegraphics[width=0.99\textwidth, height=0.7\textwidth]{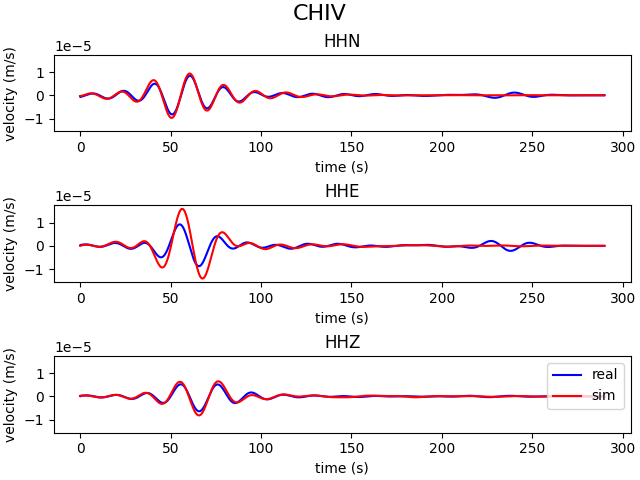}
     \end{subfigure}
     \begin{subfigure}[b]{0.24\textwidth}
         \centering
         \includegraphics[width=0.99\textwidth, height=0.7\textwidth]{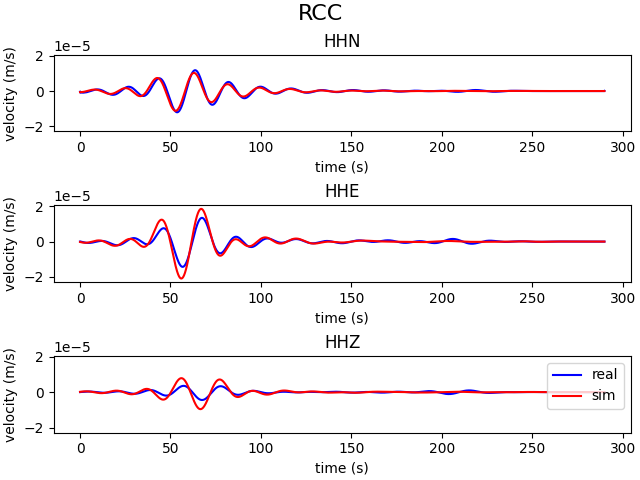}
     \end{subfigure}
     \begin{subfigure}[b]{0.24\textwidth}
         \centering
         \includegraphics[width=0.99\textwidth, height=0.7\textwidth]{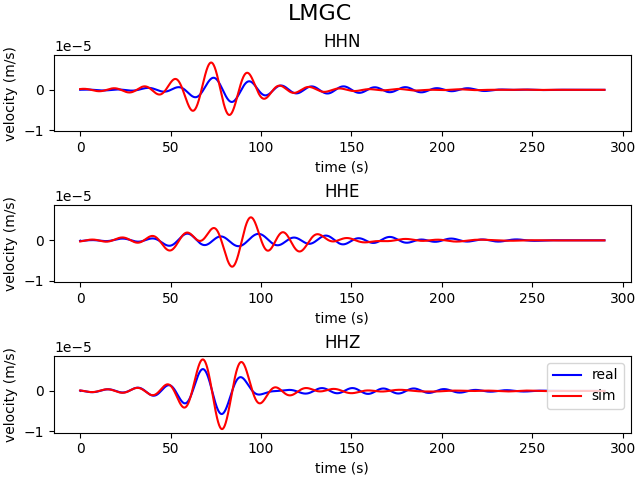}
     \end{subfigure}
     \begin{subfigure}[b]{0.24\textwidth}
         \centering
         \includegraphics[width=0.99\textwidth, height=0.7\textwidth]{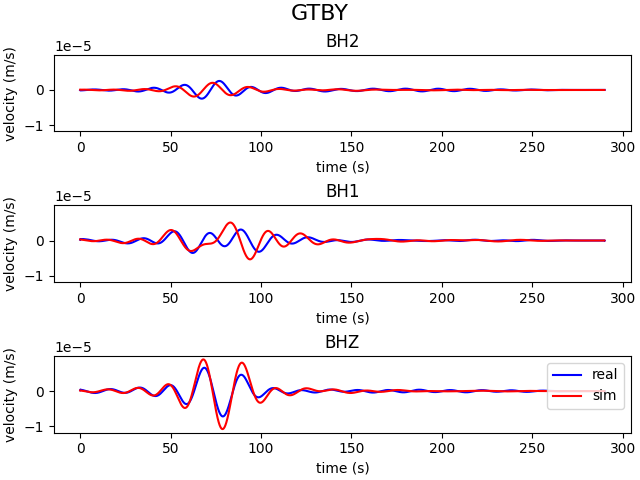}
     \end{subfigure}
     \begin{subfigure}[b]{0.24\textwidth}
         \centering
         \includegraphics[width=0.99\textwidth, height=0.7\textwidth]{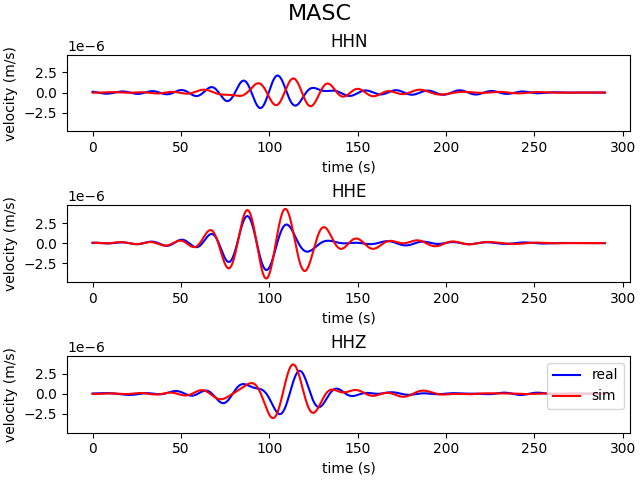}
     \end{subfigure}
     \begin{subfigure}[b]{0.24\textwidth}
         \centering
         \includegraphics[width=0.99\textwidth, height=0.7\textwidth]{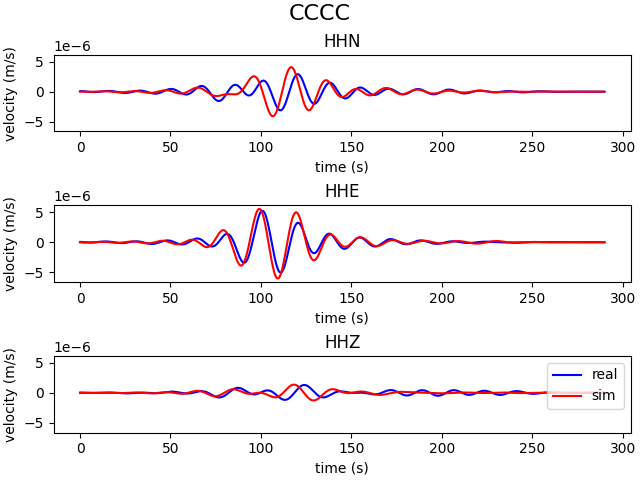}
     \end{subfigure}
     \begin{subfigure}[b]{0.24\textwidth}
         \centering
         \includegraphics[width=0.99\textwidth, height=0.7\textwidth]{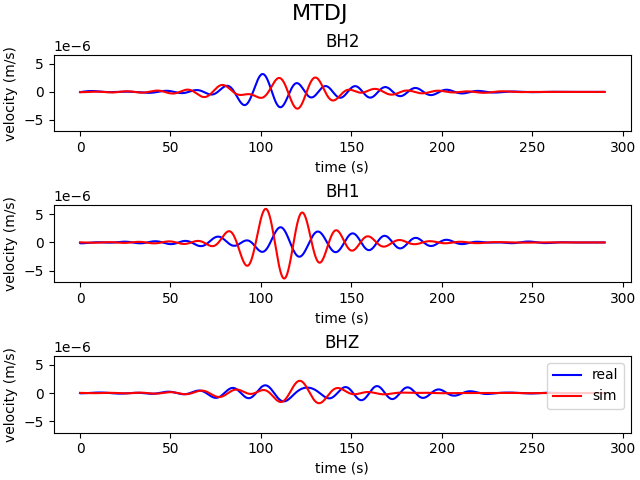}
     \end{subfigure}
     \begin{subfigure}[b]{0.24\textwidth}
         \centering
         \includegraphics[width=0.99\textwidth, height=0.7\textwidth]{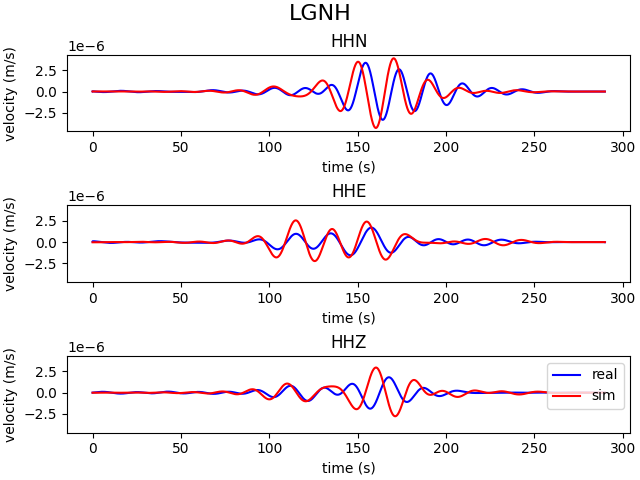}
     \end{subfigure}
     \caption{Comparison of simulated station seismograms (in red) against real seismograms (in blue) for a level of detail 3. We employed a band-pass filter with corner frequencies of 0.02 and 0.06 Hz}
     \label{fig:results_lvl_2}
\end{figure}

In figure \ref{fig:results_lvl_2} we compare the resulting simulated seismogram readings (in red) against the real seismogram (in blue) for level of detail 3. With this level is possible to simulate sources with frequencies of only up to 0.075 Hz. In order to compare our simulated seismograms against real ones, we employed a band-pass filter in both the simulated and real seismograms. The bans-pass filter employed had corner frequencies of 0.02 and 0.06 Hz. We can observe that the simulated seismograms are generally in accordance with the real seismograms, with discrepancies arising the farther away the station is with respect to the earthquake centroid. The medium parameters, taken from \cite{medium_params}, are more accurate in the vicinity of the centroid of the earthquake used in this experiment. Then, it is natural that the velocity model near the centroid of the earthquake is more accurate, and as a consequence the stations nearer to it are simulated more accurately.


\begin{figure}[h]
     \centering  
     \begin{subfigure}[b]{0.48\textwidth}
         \centering
         \includegraphics[width=0.99\textwidth, height=0.7\textwidth]{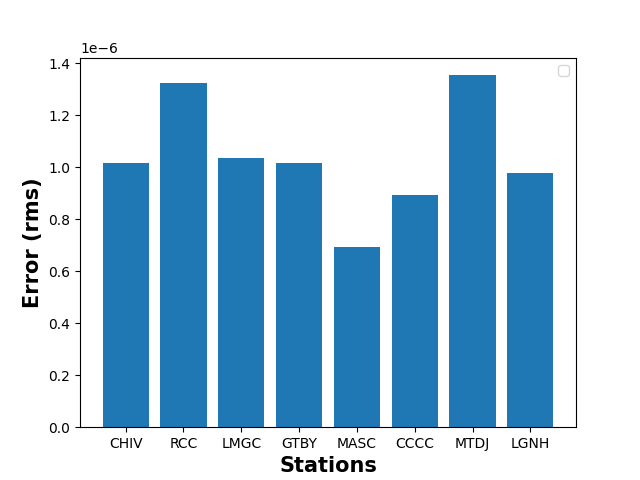}
     \end{subfigure}
     \begin{subfigure}[b]{0.48\textwidth}
         \centering
         \includegraphics[width=0.99\textwidth, height=0.7\textwidth]{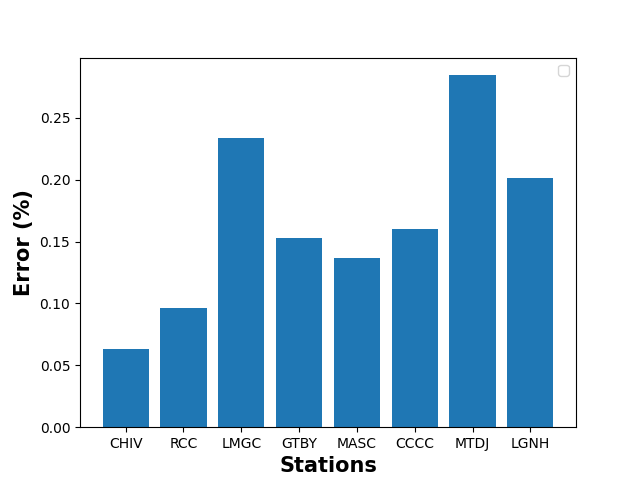}
     \end{subfigure}
     \caption{left: rms error for each station. right: relative error}
     \label{fig:error}
\end{figure}

In figure \ref{fig:error}, the graph of the left depicts the rms error between the simulated seismogram and the real seismogram, for each station in the network. We can see that the rms error is approximately the same for all stations. This is a consequence of the inversion of the moment tensor made with ISOLA. As ISOLAs inversions method tries to minimize the sum of the rms errors for all stations, it is to be expected that a minimum in the inversion would be reached when all stations have a low and similar rms error.

The graph on the right of figure \ref{fig:error} shows the mean relative error for each station. In this graph we can see that the relative error is lower in stations nearer to the earthquakes centroid. This is so because the medium parameters utilized are more accurate near the earthquakes centroid, as was noted before.

\subsection{Speed-up results}

We tested the system in a budget notebook HP probook 445 G7, with a AMD APU composed of a Ryzen 7 4700U CPU with a Vega-based GPU. The notebook has 24Gb DDR4 of RAM.

\begin{figure}[h]
     \centering  
     \begin{subfigure}[b]{0.48\textwidth}
         \centering
         \includegraphics[width=0.99\textwidth, height=0.7\textwidth]{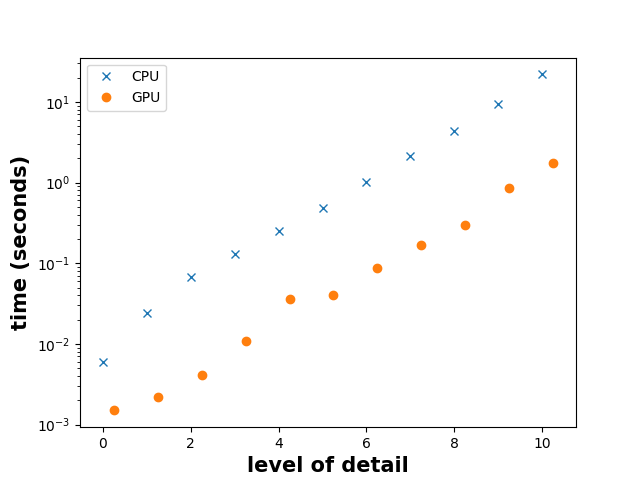}
     \end{subfigure}
     \begin{subfigure}[b]{0.48\textwidth}
         \centering
         \includegraphics[width=0.99\textwidth, height=0.7\textwidth]{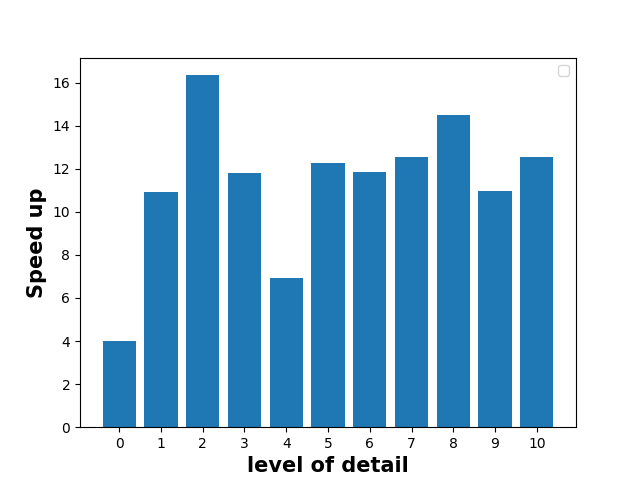}
     \end{subfigure}
     \caption{left: time to compute one time step, for the GPU and CPU implementation, for each level of detail. right: GPUs speed up for each level of detail}
     \label{fig:times}
\end{figure}

We compared the simulation step time of the CPU and GPU implementations, for each level of detail used. The results are summarized in figure \ref{fig:times} 


The mean speed up of our GPU implementation compared with the naive CPU solution is of a factor of 12.0. This is the common speed-up found in the literature \cite{Simulacion_usando_GPU_1}. This shows that our implementation is capable of providing good speed-ups compared with a system running on a CPU. Its worth mentioning that it is possible that our naive CPU implementation may not be well optimized for this workload. Even though openblas is a well known and widely utilized CPU linear algebra library, we observe that during the experiments the CPU utilization is just around 30\%. As elastic wavefield simulation is an memory-bound algorithm \cite{Simulacion_usando_GPU_1}, improvements in the CPU implementation could be made by optimizing memory access. But such optimization is not needed in our GLSL implementation, as memory-bound algorithms are common in 3D graphic workloads, and OpenGL is already optimized for such computing tasks.

We can also observe that, for the lowest level of detail, the speed up achieved is only of a factor of 4.0. This is also an expected result, as the advantages of utilizing GPUs are better realized when the workload consist of larger quantities of parallel computations.

\subsection{Real-usage usefulness}

Our ultimate objective is to allow the usage of low end consumer level hardware to accomplish full waveform inversion methods to recover detailed models of medium parameters. To assess the usefulness of our method to reach that goal, we analyzed the total simulation time. As can be seen in table \ref{tab:lvl}, not only the step simulation time increased with each successive increase of level of detail, but the total time steps as well.
Invertion methods often require many simulations to perform one iteration, and several iterations for the algorithm to converge. For this reason, it is more significant the total time required for the simulations rather than the time taken for 1 simulation step.
To test the total time, we imagine an scenario where 1000 simulations are needed per iteration, and 10 iterations are required for the algorithm to converge to a solution.

A summary of the total time needed to accomplish one simulation is shown in table \ref{tab:simulation_times_2}

\begin{table}[ht]
\centering
\caption{Simulation times}
\begin{tabular}[t]{|c|c|c|c|}
\hline
lvl & CPU total time & GPU total time (dd:hh:mm:ss)\\
\hline
0 & 00:00:00:10 & 00:00:00:03 \\
\hline
1 & 00:00:00:40 & 00:00:00:04 \\
\hline
2 & 00:00:01:53 & 00:00:00:07 \\
\hline
3 & 00:00:03:41 & 00:00:00:18 \\
\hline
4 & 00:00:14:10 & 00:00:02:02 \\
\hline
5 & 00:00:27:46 & 00:00:02:16 \\
\hline
6 & 00:00:58:22 & 00:00:04:55 \\
\hline
7 & 00:10:03:30 & 00:00:48:10 \\
\hline
8 & 00:20:32:30 & 00:01:25:00 \\
\hline
9 & 01:19:57:50 & 00:04:00:50  \\
\hline
10 & 04:07:27:50 & 00:08:15:50 \\
\hline
\end{tabular}
\label{tab:simulation_times_2}
\end{table}

For a level of detail of 3, we can perform 1000 simulation in about 5 hours using our GPU implementation. If the algorithm takes 10 iterations to converge, that adds to a little above 2 days to perform the inversion. For level of detail 4, our GPU implementation would take 33 hours to complete 1000 simulations, and to reach 10 iterations 13 days of computing time would be needed. And for level of detail 7, our system would take 150 days to complete 10 iterations.

Clearly, mid-tier notebooks are not suited to perform high level of detail seismic inversion in a reasonable amount of time. Nevertheless, our system does allow to complete reasonable levels of detail, with a grid size of 128x128x64 and a maximum source frequency of 0.075 Hz, within a time frame of 2 days. This could be use useful to accomplish coarse inversions with readily available hardware, and then use those result to initialize with a near-correct model a high level of detail inversion in a high performance computer cluster.

\section{Conclusions}
Our implementation is capable of providing speed-ups of a factor of 12, compared  with  a  naive CPU implementation. This number is probably an over estimation, as our CPU implementation may not be well optimized for this workload. Nevertheless, our GLSL implementation took relatively minimum programming efforts, as neither PU workload distribution nor memory access had to be manually programmed, all those task were accomplished by the OpenGL driver. Also, as elastic wavefield simulation is a memory-bound algorithm, it is to be expected than GPU implementations outperform CPU implementation, as the former has a much larger memory bandwidth.

We benchmarked our implementation in a hypothetical real inversion scenario, computing the total time it would take a budget notebook to accomplish 10 iterations of 1000 simulations. Our results show that even when is not possible to achieve high grid size inversions using this kind of low performance hardware, it is able to accomplish mid-range grid sizes, of about 128x128x64, in a reasonable amount of time. This could be used to get coarse-level inversion with readily available hardware, and afterwards use those results to initialize a high level of detail inversion in a high performance computer cluster.

\section{Acknowledgements}
We would like to thank to Centro Nacional de Investigaciones Sismológicas of Del Ministerio de Ciencia Tecnología y Medio ambiente from Cuba as well as the IRIS Data Management Center (USA) for the waveforms obtained. Our research was founded PICT2019- 2019- 00854, PROIPRO-UNSL 03-1520 and EU35-UNSL10806 proyects.
 
\section{Computer Code Availability}
The code is available at: \emph{  https://github.com/nosemeocurreapodo/GLSL-Elastic-3D-Wavefield-Simulation.git} under the GPL licence. It is written in the GLSL shading languaje, with function calls from a python script. The hardware requirements is a computer with a GPU with GLSL 3.3 capabilities.


\begin{thebibliography}{20}

\bibitem[Robertsson and Blanch 2011]{paper_finite_difference}
Robertsson J.O.A. and Blanch J.O. "Numerical Methods, Finite Difference", Encyclopedia of Solid Earth Geophysics. Encyclopedia of Earth Sciences Series. 2011 Springer, Dordrecht. https://doi.org/10.1007/978-90-481-8702-7

\bibitem[Virieux 1984]{paper_virieux_1}
Virieux, Jean. "P-SV wave propagation in heterogeneous media: Velocity-stress finite-difference method". Geophysics 51. 1984 889-901. 10.1190/1.1442147. 

\bibitem[Virieux 1986]{paper_virieux_2}
Jean Virieux, "P-SV wave propagation in heterogeneous media: Velocity‐stress finite‐difference method", GEOPHYSICS 1986 51, 889-901 

\bibitem[Sokos and Zahradnik 2008]{paper_ISOLA}
Sokos, E. N., Zahradnik, J. "ISOLA a Fortran code and a Matlab GUI to perform multiple-point source inversion of seismic data", Computers \& Geosciences, Volume 34, Issue 8, August 2008, Pages 967-977, ISSN 0098-3004, DOI: 10.1016/j.cageo.2007.07.005.

\bibitem[Sokos and Zahradnik 2013]{paper_ISOLA_2}
Sokos, E. and Zahradník, J. "Evaluating Centroid‐Moment‐Tensor Uncertainty in the New Version of ISOLA Software", Seismological Research Letters, July/August 2013, v. 84, p. 656-665, doi:10.1785/0220130002

\bibitem[Virieux et al. 2011]{review_de_metodos}
Virieux, J., Calandra, H. and Plessix, R.‐É. (2011), A review of the spectral, pseudo‐spectral, finite‐difference and finite‐element modelling techniques for geophysical imaging. Geophysical Prospecting, 59: 794-813. https://doi.org/10.1111/j.1365-2478.2011.00967.x

\bibitem[Kikuchi and Kanamori 1982]{kikuchi_1}
Masayuki Kikuchi, Hiroo Kanamori, "Inversion of complex body waves", Bulletin of the Seismological Society of America 1982; 72 (2): 491–506. 

\bibitem[Kikuchi and Kanamori 1991]{kikuchi_2}
Masayuki Kikuchi, Hiroo Kanamori, "Inversion of complex body waves—III". Bulletin of the Seismological Society of America 1991; 81 (6): 2335–2350. 

\bibitem[Herrmann 2013]{Herrmann}
Herrmann, R. B. "Computer programs in seismology: An evolving tool for instruction and research", Seismic Research Letters 2013 84, 1081-1088, doi:10.1785/0220110096

\bibitem[Kadlubiak et al. 2018]{Simulacion_usando_GPU_1}
Kadlubiak, K; Jaros, J; Treeby, BE; GPU-Accelerated Simulation of Elastic Wave Propagation. In: Smari, WW and Zinedine, K, (eds.) Proceedings of 2018 International Conference on High Performance Computing \& Simulation (HPCS). (pp. pp. 188-195). IEEE: Orleans, France. 
 
\bibitem[Kazei et al. 2017]{Simulacion_usando_GPU_2}
Kazei, Vladimir; Masmoudi, N.; Oh, J-W; Tzivanakis, Christos; Alkhalifah, Tariq. (2017). From CPU to GPU in Two Days: 3D Elastic Orthorhombic Modeling with OpenAcc.. 10.3997/2214-4609.201702323.
 
\bibitem[Weiss et al. 2013]{Simulacion_usando_GPU_3}
Weiss, Robin; Shragge, Jeffrey. (2013). Solving 3D anisotropic elastic wave equations on parallel GPU devices. Geophysics. 78. 7-F15. 10.1190/geo2012-0063.1. 

\bibitem[You et al. 2013]{Simulacion_usando_GPU_4}
Y. You et al., "Accelerating the 3D Elastic Wave Forward Modeling on GPU and MIC," 2013 IEEE International Symposium on Parallel \& Distributed Processing, Workshops and Phd Forum, Cambridge, MA, USA, 2013, pp. 1088-1096, doi: 10.1109/IPDPSW.2013.216.

\bibitem[Graves 1996]{Finite_difference_graves}
Robert W. Graves; Simulating seismic wave propagation in 3D elastic media using staggered-grid finite differences. Bulletin of the Seismological Society of America 1996;; 86 (4): 1091–1106. doi:



\bibitem[Perez-Ruiz 2005]{superficie_1}
Perez-Ruiz, J.. (2005). Simulation of an Irregular Free Surface with a Displacement Finite-Difference Scheme. Bulletin of The Seismological Society of America - BULL SEISMOL SOC AMER. 95. 2216-2231. 10.1785/0120050014. 

\bibitem[Robertsson 1996]{superficie_2}
Johan O. A. Robertsson, A numerical free‐surface condition for elastic/viscoelastic finite‐difference modeling in the presence of topography, GEOPHYSICS 1996 61:6, 1921-1934 

\bibitem[Zahradnik et al. 1993]{superficie_3}
Jir˘í Zahradník, Peter Moczo, Frantis˘ek Hron; Testing four elastic finite-difference schemes for behavior at discontinuities. Bulletin of the Seismological Society of America 1993;; 83 (1): 107–129. doi: 
\bibitem[Barfoot 2017]{Barfoot}
Barfoot, T. (2017). State Estimation for Robotics. Cambridge: Cambridge University Press. doi:10.1017/9781316671528

\bibitem[Talwani et al. 1959]{relacion_rho}
Talwani, M., Sutton, G.H., Worzel, J.L., 1959. A crustal section across the Puerto Rico trench. J. Geophys. Res. 64, 1545–1555. doi:10.1029/JZ064i010p01545


\bibitem[Moreno et al. 2002]{medium_params}
Bladimir Moreno, Margaret Grandison, Kuvvet Atakan, Crustal velocity model along the southern Cuban margin: implications for the tectonic regime at an active plate boundary, Geophysical Journal International, Volume 151, Issue 2, November 2002, Pages 632–645, https://doi.org/10.1046/j.1365-246X.2002.01810.x

\end{thebibliography}
\end{document}